\pdfoutput=1
\documentclass[11pt]{article}
\usepackage{jheppub}

\usepackage{customprelude}

\title{\centering Exceptional $\cN=3$ theories}

\author{I\~naki Garc\'ia-Etxebarria$^{a}$}
\author{and Diego Regalado$^{a,b}$}

\affiliation{$^{a}$Max Planck Institute for Physics, F\"ohringer Ring 6, 80805 Munich, Germany}
\affiliation{$^{b}$Institute for Theoretical Physics and
Center for Extreme Matter and Emergent Phenomena,\\
Utrecht University, Leuvenlaan 4, 3584 CE Utrecht, The Netherlands}

\emailAdd{inaki@mpp.mpg.de}
\emailAdd{regalado@mpp.mpg.de}

\abstract{We present a new construction of four dimensional $\cN=3$
  theories, given by M5 branes wrapping a $T^2$ in an M-theory U-fold
  background. The resulting setup generalizes the one used in the
  usual class $\cS$ construction of four dimensional theories by using
  an extra discrete symmetry on the M5 worldvolume. Together with the
  M-theory U-fold description of $(0,2)$ $E$-type six-dimensional
  SCFTs, this allows to construct new, exceptional, $\cN=3$ theories.}

\begin{document}

\makeatletter
\let\old@fpheader\@fpheader
\renewcommand{\@fpheader}{\old@fpheader\hfill
MPP-2016-328}
\makeatother

\maketitle

\section{Introduction}

One of the biggest strengths of string theory lies on its ability to
reformulate field theory questions in terms of geometry. In particular
cases the relevant geometry is of a particularly tractable form (for
instance, a Calabi-Yau manifold), and we can apply powerful techniques
in algebraic geometry to study various aspects of the associated field
theory. Very well known examples are the Seiberg-Witten solutions of
$\cN=2$ SYM and SQCD with gauge group $SU(2)$
\cite{Seiberg:1994rs,Seiberg:1994aj}, which are beautifully geometrized
in string theory in terms of F-theory \cite{Banks:1996nj}. More
generally, a large number of field theory results can be understood in
terms of ``geometric engineering'' of QFTs \cite{Katz:1996fh}, which
reduces subtle questions in field theory to questions about string
theory on specific geometries.

Along somewhat related lines, starting in particular with
\cite{Gaiotto:2009we}, it has been recently realized that for a
particular class of $\cN=2$ theories, namely those obtainable from
compactification of the six dimensional $(0,2)$ SCFT on a Riemann
surface, much of the interesting information of the four dimensional
theory can be understood in terms of properties of the
compactification space. The resulting formalism is very rich and
powerful, and has yielded beautiful insights into the properties of
four dimensional field theories. (See \cite{Tachikawa:2013kta} for a
nice review of some of these developments.)

Nevertheless, it is well known that classical geometry is not the only
context in which string theory is well defined, so a natural question
to ask is whether the ideas above can be extended in an interesting
way once we allow ourselves to abandon the realm of classical
geometry. A particular case of interest to us here is that of
U-manifolds
\cite{Kumar:1996zx,Hellerman:2002ax,Hull:2004in,Hull:2007zu}: these
are spaces which are locally geometric, but which involve string
dualities in the transition functions between local patches. Clearly,
ordinary geometries are a subclass of such constructions, where the
patching functions are diffeomorphisms, but one can reasonably expect
the class of non-geometric constructions to be significantly larger
than the class of constructions with a geometric interpretation. If
this expectation holds, it is then also reasonable to expect that the
space of field theories accessible using non-geometric techniques is
also significantly larger than that accessible using ordinary
geometric constructions.

In this note we aim to give some first steps in this direction, by
constructing a class of non-geometric compactifications of M-theory
which engineer various interesting field theories. The theories that
we construct explicitly in this note are
\begin{itemize}
\item The
  $\cN=3$ theories constructed in \cite{Garcia-Etxebarria:2015wns},
  rewritten as M5 branes on a
  $T^2$ inside a non-geometric compactification.
\item The six-dimensional
  $(0,2)$ SCFTs of exceptional type arising directly from a
  non-geometric compactification of M-theory down to six dimensions.
\item A new class of
  $\cN=3$ theories associated to the exceptional
  $(0,2)$ theories by compactification on a
  $T^2$ inside a non-geometric background, combining the two
  constructions above.
\end{itemize}

The first two classes of theories are already known from geometric
constructions, but the third one is new. One can already find evidence
for its existence from the four dimensional field theory perspective.
Indeed, the construction of the $\cN=3$ theories presented in
\cite{Garcia-Etxebarria:2015wns}, as quotients of $\cN=4$ $U(N)$ SYM,
relies on having an R-symmetry group $SO(6)_R$ together with an
enhanced symmetry (for certain values of the coupling) contained in
the duality group $SL(2,\bZ)$. These properties are not exclusive of
the $U(N)$ theory, but are also present in $\cN=4$ SYM with gauge
group $D_N$ or $E_n$, which are self-dual under Montonen-Olive
duality.\footnote{\label{fn:exceptional-duality}The action of duality
  for non-simply laced groups is more complicated, and in particular
  it also acts on the moduli space for the $G_2$ and $F_4$ theories
  \cite{Argyres:2006qr}. It would be rather interesting to extend the
  formalism in this paper to these cases.}$^,$\footnote{More
  precisely, in the cases of $E_6$ and $E_7$ the self-dual forms of
  the group that will appear in our geometric construction are
  $(E_6\times U(1))/\bZ_3$ and $(E_7\times U(1))/\bZ_2$, with the
  extra $U(1)$ factors associated to center-of-mass modes of the
  string configuration. This is in analogy with the fact that a stack
  of M5 branes on $T^2$ gives rise to a gauge group
  $U(N)=(SU(N)\times U(1))/\bZ_N$, and not simply $SU(N)$. The
  difference is important since neither $E_6$ or $E_7$ are invariant
  under S-duality, but rather map to their adjoint forms $E_6/\bZ_3$
  and $E_7/\bZ_2$.} Thus, it is natural to assume that one can take an
$\cN=3$ quotient of these $\cN=4$ theories. In this note we will focus
on the exceptional cases, the generalization to the orthogonal case
being straightforward.

We will provide an M-theory construction of these theories in terms of
singular U-folds, which we expect to be intrinsically non-geometric.
What we mean by ``intrinsically non-geometric'' is simply that there
is no duality frame in which the system is described by string theory
in a geometric background.\footnote{Our expectation is based on the
  fact that the two ingredients we combine for constructing these two
  theories are geometric in distinct duality frames, or more
  concretely because we take an $O(5,5;\bZ)$ U-duality action which
  cannot be conjugated into a geometric subgroup. This shows that our
  construction cannot be conjugated into pure geometry, but it does
  not show that a disconnected class of geometric constructions for
  these exceptional $\cN=3$ theories cannot exist.}  This does not
mean that geometry is entirely useless: as we shall see some aspects
of the problem can still be fruitfully geometrized using arguments
similar to those in \cite{Kumar:1996zx} and \cite{Martucci:2012jk,
  Braun:2013yla, Candelas:2014jma, Candelas:2014kma}.

\medskip

We will start in \S\ref{sec:known-N=3} by reconstructing the $\cN=3$
theories found in \cite{Garcia-Etxebarria:2015wns} in terms of an M5
wrapping a $T^2$ in an M-theory U-fold background. In \S\ref{sec:ADE}
we will construct the $E$-type $(0,2)$ theories in six dimensions in
terms of M-theory compactified on a five-manifold (elaborating on an
observation in \cite{Hellerman:2002ax,Font:2016odl}). We then combine
both constructions in \S\ref{sec:exceptional-N=3} in order to engineer
the theories of interest. We conclude in \S\ref{sec:conclusions} with
a discussion of the (numerous) directions for further research.

\section{M5 brane construction of known $\cN=3$ theories}
\label{sec:known-N=3}

In this section we obtain the four dimensional $\cN=3$ theories
constructed in \cite{Garcia-Etxebarria:2015wns} (see also
\cite{Ferrara:1998zt,Aharony:2015oyb,Nishinaka:2016hbw,Argyres:2016xua,Aharony:2016kai,Imamura:2016abe})
from the $(0,2)$ $A_{N-1}$ theory in six dimensions.\footnote{Strictly
  speaking, the torus compactification of the $(2,0)$ $A_{N-1}$ theory
  (as engineered by putting IIB string theory on $\bC^2/\bZ_N$, for
  example) yields a $\cN=4$ SYM theory with algebra
  $\mathfrak{su}(N)$, and thus a gauge group such as $SU(N)$ or
  $SU(N)/\bZ_N$. In the rest of the paper, when we talk about the
  $(2,0)$ $A_{N-1}$ theory, we actually mean the six dimensional
  $(2,0)$ theory living on a stack of M5 branes, whose torus
  compactification yields $\cN=4$ $U(N)$ SYM. We will never make use
  of the ``genuine'' $A_{N-1}$ theory, so hopefully no confusion will
  arise from our imprecise use of language.} Along the way we will be
naturally lead to consider non-geometric compactifications of
M-theory.

\subsection{S-fold construction}
\label{sec:Sfold}

Let us start by briefly reviewing the construction of the four
dimensional $\cN=3$ theories of \cite{Garcia-Etxebarria:2015wns}. The
basic idea is to take certain quotient of four dimensional $\cN=4$ SYM
with gauge group $U(N)$ by a $\bZ^{\cN=3}_k$ symmetry of the theory,
which includes both R-symmetry rotations and $SL(2,\bZ)$ duality. In
order for the quotient to make sense, the coupling constant of the
original theory must be tuned to a particular value that lies at
strong coupling, so that part of the duality group becomes an actual
symmetry. The quotient then projects out the corresponding marginal
deformation together with four of the supercharges
\cite{Aharony:2015oyb,Cordova:2016xhm}. More specifically, the
quotient we need to take is $\bZ^{\cN=3}_k=\bZ_k^R\cdot \bZ_k^{\tau}$
for $k=3,4,6$, where $\bZ_k^R$ is generated by
\begin{equation}
R_k=\left ( \begin{array}{ccc} \hat R_k^{-1}&0&0\\
0&\hat R_k&0\\
0&0& \hat R_k
\end{array}\right )\in SO(6)_R
\end{equation}
with $\hat R_k$ being a $2\pi/k$ rotation in two dimensions. The S-duality quotient $\bZ_k^{\tau}$ is generated, for $k=3,4,6$, by the following matrices in $SL(2,\bZ)$
\begin{equation}
S_3=\left (\begin{array}{cc} 0&-1\\1&-1\end{array}\right ),\qquad S_4=\left (\begin{array}{cc} 0&-1\\1&0\end{array}\right ),\qquad S_6=\left (\begin{array}{cc} 1&-1\\1&0\end{array}\right )\,.
\end{equation}

A simple string construction of these theories is to consider M-theory
with $N$ M2 branes in $\underline{\bR^{1,2}}\times \bC^3\times T^2$,
where the underline denotes the dimensions spanned by the M2
branes. Upon taking the F-theory limit, the M2 branes lift to D3 branes in
$\underline{\bR^{1,3}}\times \bC^3$, which realizes four dimensional
$\cN=4$ SYM with gauge group $U(N)$ in their worldvolume. The key
point is that in the M-theory description, both the $SO(6)_R$
R-symmetry group and the $SL(2,\bZ)$ duality are manifest
geometrically. Thus, one can take a conventional orbifold in the
M-theory side, namely
$\underline{\bR^{1,2}}\times (\bC^3\times T^2)/\bZ_k$, which after
taking the F-theory limit reproduces the quotient in the
four-dimensional gauge theory. This yields an $\cN=3$ theory on the
worldvolume of a stack of D3 branes probing a generalized orientifold,
dubbed S-fold in \cite{Aharony:2016kai}.\footnote{We restrict to the
  case without torsion fluxes in the whole paper. A systematic
  discussion of fluxes in S-folds can be found in
  \cite{Aharony:2016kai}. As we discuss in the conclusions, we expect
  that some or all of these discrete degrees of freedom can be encoded
  into a discrete twist along the compactification torus.}

\subsection{$\cN=3$ from six dimensions}
\label{sec:sixdim}

Now we would like to obtain the four dimensional $\cN=3$ theories from
the $(0,2)$ $A_{N-1}$ superconformal field theory in six
dimensions. The R-symmetry group of these theories is $SO(5)_R$ and
the supercharges transform in the $(\mathbf 4,\mathbf 4)$ of
$SO(5,1)\times SO(5)_R$.

\subsubsection{From M5 branes to D3 branes}

Four dimensional $\cN=4$ SYM with gauge group $U(N)$ can be obtained
from the $(0,2)$ $A_{N-1}$ theory by compactifying it on a torus
\cite{Witten:1995zh}. In this construction, the S-duality group
$SL(2,\bZ)$ of the four dimensional theory is manifest geometrically,
as the group of large diffeomorphisms acting on the torus. In
contrast, the full R-symmetry group $SO(6)_R$ is only present in the
limit in which the size of the torus vanishes, where $SO(5)_R$
enhances to $SO(6)_R$. Since the quotient we want to take in the
$\cN=4$ theory involves a subgroup of $SO(6)_R$ which is not in
$SO(5)_R$, it is not immediately clear how to proceed in terms of the
$(0,2)$ theory when the size of the torus is finite. In order to do
so, we need to make the symmetry we want to quotient by manifest in
the UV. In the following we do so by starting with a specific M-theory
configuration of M5 branes and interpret the result in field theory
terms afterwards.

Let us consider a system of $N$ M5 branes on $\underline{\bR^{1,3}\times S^1_M\times S^1_T}\times  S^1_E\times \bC^2$. If we reduce along the M-theory circle $S^1_M$, we have Type IIA with $N$ D4 branes on $\underline{\bR^{1,3}\times S^1_T} \times S^1_E\times \bC^2$. T-duality along $S^1_T$ brings us to Type IIB with $N$ D3 branes on $\underline{\bR^{1,3}}\times \tilde S^1_T\times  S^1_E\times \bC^2$. In this last picture we see that, due to the two circles in the transverse space to the D3 branes, the R-symmetry group $SO(6)_R$ is broken, and is only recovered in the IR where these circles decompactify, as mentioned earlier.  In general, we expect it to be broken to $SO(4)\times \bZ_2$, where $SO(4)$ acts on $\bC^2$ and $\bZ_2$ acts by reflection on the torus $\tilde T_E^2=\tilde S^1_T\times S^1_E$. However, if we tune the complex structure $\tilde \tau_E$ of $\tilde T^2_E$ to be, for example, $\tilde \tau_E=i$, we see that the symmetry group is enhanced to $SO(4)\times \bZ_4$, with $\bZ_4$ acting on $\tilde T^2_E$ as a rotation by $\pi/2$, namely
\begin{equation}\label{z4}
\bZ_4\ :\ (\tilde x_T,x_E)\longrightarrow (-x_E,\tilde x_T)
\end{equation}
where $(\tilde x_T,x_E)$ are coordinates on $\tilde T^2_E$. Thus, in
that particular situation, even though we do not have the full
R-symmetry $SO(6)_R$, we have precisely the $\bZ_4^R$ within $SO(6)_R$
that we need in the construction of the $\cN=3$ theories. Similarly,
if we tune $\tilde \tau_E=e^{i\pi/3}$, we have an enhanced $\bZ_6$
R-symmetry, which allows to construct the rest of the $\cN=3$ theories
in \cite{Garcia-Etxebarria:2015wns}. In the following we restrict to
the case $\bZ_4$ for simplicity, with generalization to the other
cases being straightforward.

Since the original description in terms of M5 branes is dual to the one involving D3 branes, we know that such $\bZ_4$ symmetry must be there too. In order to identify it in terms of M5 branes, we dualize back tracing carefully the $\bZ_4$ that acts on the torus $\tilde T^2_E$.

\subsubsection{From D3 branes back to M5 branes}

The first step is to T-dualize along $\tilde S^1_T$, which gives Type IIA with $N$ D4 branes on $\underline{\bR^{1,3}\times S^1_T} \times S^1_E\times \bC^2$. Since upon T-duality we exchange $\tau\leftrightarrow \rho$, the complexified K\"ahler parameter
\begin{align}
\rho=\int_{T^2_E} B+i\sqrt{\det G}
\end{align}
of $T^2_E=S^1_T\times S^1_E$ is equal to $\rho_E=i$. Here $G$ is the metric on the torus $T^2_E$. In particular, this means that the radii of $T^2_E$ (in string frame) are constrained by $r_Tr_E=1$ and that the NSNS B-field is zero when integrated over $T^2_E$.

In this picture, the $\bZ_4$ symmetry \eqref{z4} is no longer geometric since, for the closed string sector, it exchanges momentum states in one direction with winding modes in the other. Type IIA on a two-torus has a T-duality group $O(2,2;\bZ)$, which can be written as
\begin{equation}\label{Tgroup}
O(2,2;\bZ)=(SL(2,\bZ)_{\tau_E} \times SL(2,\bZ)_{\rho_E})\rtimes
(\bZ_2^{\tau_E\leftrightarrow\rho_E} \times
\bZ_2^{(\tau_E,\rho_E)\leftrightarrow (-\ov\tau_E,-\ov\rho_E)}).
\end{equation}
Here $SL(2,\bZ)_{\tau_E}$ acts geometrically on $T^2_E$, while $SL(2,\bZ)_{\rho_E}$ does it non-geometrically, since it acts on $\rho_E$, which contains the volume modulus, by the usual linear fractional transformations. Generically, this is a duality and not a symmetry, in the sense that it identifies states with different values of the fields at infinity. However, for the special value $\rho_E=i$, we see that the $\bZ_4\subset SL(2,\bZ)_{\rho}$, acting as\,\footnote{A useful way of viewing this transformation is as a T-duality along $S^1_T$, followed by a T-duality along $S^1_E$, followed by a rotation exchanging the two coordinates.}
\begin{equation}\label{inv}
\rho_E\longrightarrow -\frac{1}{\rho_E}\,,
\end{equation}
becomes a symmetry, since it leaves the asymptotic value of all the
fields fixed. Thus, for $\rho_E=i$, we may take the quotient by
$\bZ_4$. Regarding the action on the open string degrees of freedom of the D4 branes, at the massless level, this $\bZ_4$ exchanges the position along the transverse circle $S^1_E$ with the Wilson line along $S^1_T$.

\medskip

The next step is to take the M-theory lift of this configuration and, in particular, of the $\bZ_4$ action \eqref{inv}. 
As explained in \cite{Aharony:1996wp}, the duality group of M-theory on $T^3$ is given
by $SL(3,\bZ)\times SL(2,\bZ)_\rho$, where the first factor acts in
the natural way on $T^3$ and the second corresponds to certain
electric-magnetic duality in eight dimensions. Indeed, it exchanges
the M-theory three-form $C$ along the eight non-compact dimensions
with its Hodge-dual potential or, equivalently, maps unwrapped M2
branes to wrapped M5 branes. In addition, it acts on the M-theory
$\rho$ parameter by linear fractional transformations, where $\rho$ is
given now by 
\begin{align}
\rho = \int_{T^3} C+i \sqrt{\det G }
\end{align}
with $G$ the metric on $T^3$. This provides the M-theory lift of the T-duality
group of Type IIA on $T^2$.

Thus, we find that the M-theory lift consists of $N$ M5 branes on $\underline{\bR^{1,3}\times S^1_M\times S^1_T}\times  S^1_E\times \bC^2$ with $\rho=i$, where $\rho$ is now the M-theory modulus. Having $\rho=i$ implies that
\begin{equation}\label{R}
R_MR_TR_E=1\,,
\end{equation}
where $R_{M},\,R_{T},\,R_{E}$ are the radii of $S^1_M,\,S^1_T,\,S^1_E$ in the M-theory metric.

\medskip

As we mentioned earlier, the S-duality group $SL(2,\bZ)$ of four dimensional $\cN=4$ SYM arises from large diffeomorphisms of the torus wrapped by the M5 branes. Thus, in order to have $\bZ_4^\tau$ within $SL(2,\bZ)$ to be an actual symmetry, we have to tune the complex structure $\tau$ of $S^1_M\times S^1_T$ to be $\tau=i$ which, in particular, implies that 
 \begin{equation}\label{S}
 R_M=R_T\,.
 \end{equation}
Thus, requiring the presence of both the R-symmetry $\bZ_4^R$ (eq.\eqref{R}) and the S-duality $\bZ_4^\tau$ (eq.\eqref{S}) leaves only one free parameter, namely
\begin{equation}
R_M=R\,,\qquad R_T=R\,,\qquad R_E=\frac{1}{R^2}\,.
\end{equation}
Notice that the $\bZ_4^{\cN=3}$ action we want to quotient by in four dimensions to construct the $\cN=3$ theories is present for every value of $R$. The four dimensional superconformal theory is obtained when $R\rightarrow 0$. In this limit, the two-torus wrapped by the M5 branes becomes small while the transverse circle decompactifies.

\medskip

To summarize, we can obtain the four dimensional $\cN=3$ theories in this context by considering $N$ M5 branes probing certain non-geometric singularity. More explicitly, we need to consider $N$ M5 branes  on $\underline{\bR^{1,3}\times (S^1_M\times S^1_T}\times S^1_E\times \bC^2)/\bZ_k$, where $\bZ_k$ is the combined action
\begin{equation}\label{zetak}
\bZ_k=\bZ_k^R\cdot\tilde\bZ_k^R\cdot\bZ_k^\tau\,.
\end{equation} 
Here $\bZ_k^R$ is a rotation acting on $\bC^2$ generated by
\begin{equation}\label{rot3}
R_k=\left(\begin{array}{cc} \hat R_k^{-1}&0\\0&\hat R_k\end{array}\right)\,,
\end{equation}
where $\hat R_k$ is a $2\pi/k$ rotation in two dimensions. Moreover, $\tilde\bZ_k^R$ is a non-geometric quotient generated by acting on the $\rho$ parameter of $T^3=S^1_M\times S^1_T\times S^1_E$, which fixes $\rho$ to a specific value such that the volume of $T^3$ is of order one. Finally, $\bZ_k^\tau$ acts on $T^2=S^1_M\times S^1_T$ as $u\rightarrow e^{2\pi i/k}u$, where $u$ is a flat complex coordinate on $T^2$. This fixes the complex structure of the torus to a particular value.

\subsubsection{Supercharges}

It is interesting to compute the amount of supersymmetry preserved by
the configuration described above directly. If we consider $N$ M5 branes on $\underline{\bR^{1,3}\times S^1_M\times S^1_T}\times  S^1_E\times \bC^2$, the sixteen supercharges preserved by the M5 branes transform as \cite{Kumar:1996zx}\,\footnote{In our conventions, the M5 brane preserves (0,2) supersymmetry in six dimensions.}
\begin{equation}
(S_4^+,S_4^+)_{-\frac{1}{2},\frac{1}{2}}\oplus (S_4^+,S_4^-)_{-\frac{1}{2},-\frac{1}{2}}\oplus (S_4^-,S_4^+)_{\frac{1}{2},-\frac{1}{2}}\oplus (S_4^-,S_4^-)_{\frac{1}{2},\frac{1}{2}}
\end{equation}
under
\begin{equation}\label{group}
SO(1,3)\times SO(4)_R\times U(1)_\tau\times U(1)_\rho\,,
\end{equation}
where $S_4^{\pm}$ are the positive/negative chirality spinors of either $SO(1,3)$ or $SO(4)$. Furthermore, $U(1)_\tau$ corresponds to the rotations on the torus wrapped by the M5 branes and $U(1)_\rho$ is the bundle associated to the duality group $SL(2,\bZ)_\rho$, defined as follows. Given an $SL(2,\bZ)_\rho$ bundle with transition functions 
\begin{equation}\label{u1}
M=\left (\begin{array}{cc} a&b\\c&d\end{array}\right )\in SL(2,\bZ)_\rho\,,
\end{equation}
the $U(1)_\rho$ bundle is obtained by using transition functions $e^{i{\rm arg}(c\rho+d)}$.

We can compute how the supercharges transform under the discrete group \eqref{zetak} and those that are not invariant will be projected out. More explicitly, $\bZ_k^R$ acts on the supercharges as
\begin{equation}
\bZ_k^R\,:\,\begin{array}{ccc} 
(S_4^\pm,S_4^+)_{p,q}&\rightarrow& (S_4^\pm,S_4^+)_{p,q}\\
(S_4^\pm,(+\frac12,-\frac12))_{p,q}&\rightarrow& e^{-2\pi i/k}(S_4^\pm,(+\frac12,-\frac12))_{p,q}\\
(S_4^\pm,(-\frac12,+\frac12))_{p,q}&\rightarrow& e^{2\pi i/k}(S_4^\pm,(-\frac12,+\frac12))_{p,q}\,.
\end{array}
\end{equation}
Under the non-geometric action $\tilde \bZ_k^R$ we find
\begin{equation}
\tilde\bZ_k^R\,:\,\begin{array}{ccc} 
(S_4^\pm,\bullet)_{p,\frac{1}{2}}&\rightarrow& e^{\pi i/k}(S_4^\pm,\bullet)_{p,\frac{1}{2}}\\
(S_4^\pm,\bullet)_{p,-\frac{1}{2}}&\rightarrow& e^{-\pi i/k}(S_4^\pm,\bullet)_{p,-\frac{1}{2}}\,,
\end{array}
\end{equation}
where the bullets stand for omitted $S_4^\pm$ terms.
Finally, under the rotation of the torus wrapped by the M5 branes we find that
\begin{equation}
\bZ_k^\tau\,:\,\begin{array}{ccc} 
(S_4^\pm,\bullet)_{\frac{1}{2},q}&\rightarrow& e^{\pi i/k}(S_4^\pm,\bullet)_{\frac{1}{2},q}\\
(S_4^\pm,\bullet)_{-\frac{1}{2},q}&\rightarrow& e^{-\pi i/k}(S_4^\pm,\bullet)_{-\frac{1}{2},q}\,.
\end{array}
\end{equation}
Under the combined action $\bZ_k=\bZ_k^R\cdot\tilde\bZ_k^R\cdot\bZ_k^\tau$, only twelve supercharges remain invariant so we have $\cN=3$, as expected.

\subsection{Field theory interpretation}
\label{sec:field}

Up to now we have discussed how to obtain the four dimensional $\cN=3$
from six dimensions by using the M-theory construction in terms of M5
branes. However, it should be possible to understand this procedure
directly in terms of the $(0,2)$ theory. For concreteness we will
discuss the case $k=4$ in the following but the other cases work
analogously.

Let us start by looking at the moduli space of the abelian $(0,2)$
theory (just one M5 brane in flat space), which is
\begin{equation}
\mathcal M=\bR^5\,.
\end{equation}
The R-symmetry group $SO(5)_R$ acts on the moduli space in the obvious
way. Notice that there is an additional $\bZ_2$ symmetry of the theory
that acts as the element $(-1)\in O(5)$ on the moduli space and which,
in order commute with the supercharges, must act also with a minus
sign on the self-dual two-form potential $B$. In the M-theory
construction, this corresponds to the possibility of taking an M5
brane on an orbifold $\mathbb R^5/\bZ_2$. Since the resulting space is
non-orientable, the orbifold action must be accompanied by
$C\rightarrow-C$ \cite{Witten:1996md}, which induces $B\rightarrow -B$
on the M5 brane. Gauging such $\bZ_2$ provides the field theory
construction of the $(0,2)$ \mbox{$D$-type} theories in terms of the
$A$-type ones.\footnote{Strictly speaking, this provides a variant of
  the $D$-type theory in which the global $\bZ_2$ symmetry of the
  $D_N$ theory is gauged, analogous to the difference between a gauge
  theory with gauge group $SO(2N)$ and $O(2N)$. This subtlety becomes
  more clear when one looks carefully at the different possible
  boundary conditions of the holographic dual \cite{Aharony:2016kai}.}
Thus, the full symmetry group of the theory is $SO(5)_R\times \bZ_2$,
the same as the isometry group of the moduli space that leaves the
origin fixed.

When we compactify the theory on a square torus $T^2=S^1_M\times S^1_T$\,, the moduli space becomes
\begin{equation}
\mathcal M=\bR^5\times S^1_h\,,
\end{equation}
where $S^1_h$ corresponds to the scalar $\varphi$ that comes from the holonomy of the two-form potential along $T^2$, namely
\begin{equation}
\exp\left [i\int_{T^2_\tau}B\right ]=\exp\left [{iR_MR_T \, \varphi }\right ]\,,
\end{equation}
where $R_M$ and $R_T$ are the radii of the torus. From this we see that the radius of the circle $S^1_h$ in moduli space is $(R_MR_T)^{-1}$ and that this corresponds to turning on a relevant deformation in the six dimensional theory. If we compactify a direction transverse to the M5 ($\bR\rightarrow S^1_p$), the moduli space is
\begin{equation}\label{mod}
\mathcal M=\bR^4 \times S^1_p \times S^1_h
\end{equation}
with $S^1_p$ of radius $R_E$. This amounts to turning on an irrelevant deformation of the $(0,2)$ theory that breaks the R-symmetry group from $SO(5)_R$ to $SO(4)_R$, so the full symmetry group is generically $SO(4)_R\times \bZ_2$. Since the $\bZ_2$ acts on the potential as $B\rightarrow -B$, it acts by reversing all the coordinates in the moduli space \eqref{mod}.

From the M-theory construction discussed above, we expect that for a particular value of this irrelevant deformation ($R_E$), there is an enhancement of the symmetry group of the theory to $SO(4)_R\times \bZ_4$. One can find evidence for this directly in field theory by looking at the isometries of moduli space \eqref{mod}. Indeed, the two $S^1$ factors form a complex torus $T^2_{ph}=S^1_p\times S^1_h$ with complex structure $\tau_{ph}$. When $\tau_{ph}=i$, there is an enhancement of the isometry group of the moduli space to $SO(4)_R\times \bZ_4$. In particular, this happens when the radii of $S^1_p$ and $S^1_h$ are the same, namely when $R_E=(R_MR_T)^{-1}$, cf.~\eqref{R}. Thus, we conclude that the complex structure $\tau_{ph}$ is precisely the same as the $\rho$ parameter of the M-theory construction presented above. The enhanced discrete symmetry acts as
\begin{equation}
\bZ_4\ :\ (\phi,\varphi)\rightarrow(-\varphi,\phi)\,,
\end{equation}
where $\phi$ is the position along the transverse circle. We see that it exchanges the position modulus along $S^1_E$ with the holonomy coming from $B$ on $S^1_M\times S^1_T$, as expected.

In the simplest construction of the class
$\cS$ theories \cite{Gaiotto:2009we}, the transverse space to the M5
branes is taken to be
$\bR^5$. This means that the full R-symmetry group
$SO(5)_R$ can be used to topologically twist the theory and preserve
supersymmetry. In our case, by compactifying one of the directions in
the transverse space, we break the R-symmetry to
$SO(4)_R$, but for special values of the transverse radius, there is
an enhancement to $SO(4)_R\times
\bZ_4$, which is not contained in
$SO(5)_R$. This extra symmetry can be used to perform a
compactification such that the four dimensional theory preserves
$\cN=3$ supersymmetry. In this sense, our construction can be regarded
as a generalization of the class $\cS$ theories.

\section{New $\cN=3$ theories of exceptional type}

\subsection{Six dimensional (0,2) E-type theories from M-theory}
\label{sec:ADE}

The six dimensional (0,2) ADE theories first appeared as the low
energy description of ADE singularities in Type IIB
\cite{Witten:1995zh}. Such singularities are locally of the form
$\bC^2/\Gamma$, where $\Gamma$ is a finite subgroup of
$SU(2)$. However, in order to provide an M-theory realization of these
theories, it is useful to consider instead a singular elliptic
fibration over $\bC$ such that, upon decompactification of the fiber,
we recover $\bC^2/\Gamma$.\,\footnote{In our construction the transverse
  geometry has some compact directions, so the four
  dimensional theory arising upon compactification of the 6d $A_{N-1}$
  theory includes a free $U(1)$ multiplet, i.e. we have the $U(N)$
  theory.} As we go around the singularity, the fiber undergoes a
monodromy given by an element of $SL(2,\bZ)_\tau$, which characterizes
uniquely the type of singularity of the corresponding Weierstrass model (see table \ref{table:ADE}). In
particular, such a monodromy acts on the complex structure of the
fiber as
\begin{equation}
\tau\rightarrow \frac{a\tau +b}{c\tau +d}\,, \qquad \left (\begin{array}{cc} a&b\\c&d\end{array}\right )\in SL(2,\bZ)_\tau.
\end{equation}

Away from the singular point, the space is locally given by $\bC\times T^2$ and the group $SL(2,\bZ)_\tau$ is part of the T-duality group of Type IIB on $T^2$, see eq.\eqref{Tgroup}.

If we T-dualize along one of the directions of the fiber, we map Type IIB to Type IIA and exchange $\tau\leftrightarrow \rho$, which means that we have an elliptic fibration with a monodromy acting on $\rho$, not on $\tau$. This is an example of a non-geometric space, sometimes referred to as a T-fold.

As shown in table~\ref{table:ADE}, the corresponding monodromy for the case of $A_{N-1}$ acts on the K\"ahler parameter as $\rho\rightarrow\rho+N$, i.e.~a shift in the $B$-field. This means that the singularity is magnetically charged under $B$ which, together with having sixteen preserved supercharges, implies that such an object is a stack of NS5 branes. Notice that since the monodromy does not act on the volume of $T^2$, we may decompactify it.

For the case of $D_N$, the action on $\rho$ is the same as for
$A_{N-3}$, but the monodromy differs by an overall sign. This
corresponds to having $N$ NS5 branes in the presence of an ON5,
defined as Type IIA modded out by $I_4(-1)^{F_L}$
\cite{Hanany:2000fq}. 

For the exceptional cases, the interpretation is rather different. Let us consider for concreteness the case of $E_7$, for which the monodromy acts as $\rho\rightarrow -1/\rho$. Unlike the previous cases, this involves a genuine stringy duality that (for vanishing $B$-field) sends the volume of the fiber to its inverse. In particular, this implies that, at the singularity, the value of $\rho$ is given by the fixed point of $\rho\rightarrow -1/\rho$, namely $\rho=i$. Thus, we cannot decompactify the fiber, in contrast with the $A_N$ and $D_N$ cases. For $E_6$ and $E_8$, the $\rho$ modulus is fixed at the singularity to $\rho=e^{i\pi/3}$.

This shows that all the six dimensional (0,2) theories can be engineered in Type IIA, as long as we allow to have singular non-geometric compactification spaces.

\begin{table}
  \centering
  \begin{tabular}{c|c c c c c}
     & $A_{N-1}$ & $D_N$ & $E_6$ & $E_7$ & $E_8$ \\
    \hline
    Monodromy & $\left (\begin{array}{cc} 1&N\\0&1\end{array}\right )$ & $\left (\begin{array}{cc} -1&4-N\\0&-1\end{array}\right )$ & $\left (\begin{array}{cc} 0&-1\\1&-1\end{array}\right )$ & $\left (\begin{array}{cc} 0&-1\\1&0\end{array}\right )$ & $\left (\begin{array}{cc} 1&-1\\1&0\end{array}\right )$
  \end{tabular}
  \caption{Monodromy of ADE singularities in elliptic fibrations. For the case $D_N$ we restrict $N\ge 4$.}
  \label{table:ADE}
\end{table}

\medskip

Now that we have a Type IIA construction of these theories, we can
obtain the \mbox{M-theory} lift, as done in the previous section. 
The non-geometric action on the torus $T^2$ in Type IIA lifts to a 
non-geometric action on the M-theory three torus $T^3$ \cite{Aharony:1996wp}.
Thus, we find that we can engineer the
(0,2) ADE theories in M-theory by considering a non-geometric $T^3$
fibration over $\bC$, where the monodromy acts on the complexified
volume of $T^3$, as in table \ref{table:ADE}.

Similarly to the Type IIA case, we can interpret the $A_{N-1}$ and $D_N$ cases as corresponding to a stack of $N$ M5 branes, either in flat space or in an orbifold $\bR^5/\bZ_2$ \cite{Hanany:2000fq}. However, for the exceptional cases the monodromy acts non-trivially on the volume of $T^3$, so we cannot decompactify the fiber. In these cases, the six dimensional superconformal point is reached, in the IIB description, when we decompactify the fiber, namely $\rho_{\rm IIB}\rightarrow i\infty$, keeping $g_{\rm IIB}$ fixed. In terms of the M-theory data, this corresponds (for $E_7$) to taking the limit $R\rightarrow 0$ in
\begin{align}\label{scaling}
R_A=R^{-2}c^{-1}\,,\qquad R_T=R\,c^{-1}\,,\qquad R_M=R\,c^2\,,
\end{align}
where $c=(g_{IIB})^{\frac{1}{3}}$ and $R_A,\,R_T,\,R_M$ are the radii of
$T^3=S^1_A\times S^1_T\times S^1_M$ in the M-theory metric. Here
$S^1_T$ is the circle along which we T-dualize and $S^1_M$ is the
M-theory circle. Notice that the M-theory complexified volume is
$\rho=i$ for every $c$ and $R$ in \eqref{scaling}.

\medskip

Since the M-theory configuration is dual to the original Type IIB setup, we know that it preserves sixteen supercharges of the same chirality. However, it is instructive to compute this directly in M-theory for the exceptional cases. Consider for the moment M-theory on $\bR^{1,5}\times (\bC\times T^3)$, so the supercharges transform as
\begin{equation}\label{Q6d}
\left (S_6^+,\frac{1}{2},\frac{1}{2},\mathbf 2\right )\oplus\left (S_6^+,-\frac{1}{2},-\frac{1}{2},\mathbf 2\right )\oplus\left (S_6^-,\frac{1}{2},-\frac{1}{2},\mathbf 2\right )\oplus\left (S_6^-,-\frac{1}{2},\frac{1}{2},\mathbf 2\right )
\end{equation}
under $SO(1,5)\times U(1)_{\bC} \times U(1)_{\rho}\times SU(2)$, where $U(1)_\bC$ is the rotation group in $\bC$ and $S_6^{\pm}$ are positive/negative chirality spinors in six dimensions. Here $U(1)_\rho\times SU(2)$ is the maximal compact subgroup of the (continuous version of the) duality group $SL(2,\bZ)_\rho\times SL(3,\bZ)$. 

In order to compute the supercharges that survive the quotient, we may regard the non-geometric $T^3$ fibration over $\bC$ as $(\bC\times T^3)/\bZ_p$ where $\bZ_p$ acts non-geometrically. Namely, it is given by the combined action
\begin{equation}\label{zetarho}
\bZ_p=\bZ_p^\bC\cdot \bZ_p^\rho\,,
\end{equation}
where $\bZ_p^\bC$ acts on $\bC$ as a rotation by
$2\pi/p$ and $\bZ_p^\rho$ acts on
$\rho$ via the monodromy in table \ref{table:ADE}.  We have that
$p=3,4,6$ correspond to $E_6,\, E_7$ and
$E_8$, respectively. This is analogous to the statement that the
original Weierstrass model for the singular elliptic fibrations of
type $IV^*,\,III^*$ and
$II^*$ is birational to orbifolds of the form $(\bC\times
T^2)/\bZ_p$, for
$p=3,4,6$. In the geometric case, the Weierstrass model and the
orbifold are not physically equivalent, since the singularity
structure is different in each case, and thus the corresponding
superconformal field theories are also different. However, the amount
of supersymmetry preserved is indeed the same in both cases so in the
following, for simplicity, we will use an orbifold description of the
U-fold, which correctly encodes the monodromies associated to the
singularity, to compute the amount of supersymmetry preserved. As we
will see, this gives the expected number of supercharges. We stress,
however, that this is not the U-fold that yields the E-type
superconformal field theories we are interested in, which is rather
given by the U-dual of the Weierstrass model which one needs for
constructing the exceptional $(0,2)$ theories.

On the one hand, under
the rotation in $\bC$, the supercharges \eqref{Q6d} transform as
\begin{equation}\label{action1}
\bZ_p^\bC\,:\, \begin{array}{ccc}
\left (S_6^+,\pm\frac{1}{2},\pm\frac{1}{2},\mathbf 2\right )&\rightarrow& e^{\pm i\pi/p}\left (S_6^+,\pm\frac{1}{2},\pm\frac{1}{2},\mathbf 2\right )\\
\left (S_6^-,\pm\frac{1}{2},\mp\frac{1}{2},\mathbf 2\right )&\rightarrow& e^{\pm i\pi/p}\left (S_6^-,\pm\frac{1}{2},\mp\frac{1}{2},\mathbf 2\right )\,.
\end{array}
\end{equation}
On the other hand, under the non-geometric monodromy in the fiber, these transform as
\begin{equation}\label{action2}
\bZ_p^{\rho}\,:\,\begin{array}{ccc}
\left (S_6^+,\pm\frac{1}{2},\pm\frac{1}{2},\mathbf 2\right )&\rightarrow& e^{\pm i\pi/p}\left (S_6^+,\pm\frac{1}{2},\pm\frac{1}{2},\mathbf 2\right )\\
\left (S_6^-,\pm\frac{1}{2},\mp\frac{1}{2},\mathbf 2\right )&\rightarrow& e^{\mp i\pi/p}\left (S_6^-,\pm\frac{1}{2},\mp\frac{1}{2},\mathbf 2\right )\,.
\end{array}
\end{equation}
Clearly, the combined action \eqref{zetarho} preserves all the supercharges of negative chirality  and projects out the rest, which gives (0,2) supersymmetry, as expected.

\subsection{New $\cN=3$ theories}
\label{sec:exceptional-N=3}

As we mentioned in the introduction, the idea of quotienting $\cN=4$
SYM by a combination of appropriate $R$-symmetry and $SL(2,\bZ)$
transformations does not require the original group to be $U(N)$, and
should extend in particular to $\cN=4$ theories with exceptional gauge
groups, which are also self-dual for certain values of the coupling
(with the subtlety mentioned in footnote~\ref{fn:exceptional-duality}
taken into account). This field theory argument suggests that the
quotient exists and yields $\cN=3$ theories. However, with current
technology it is difficult to analyze the resulting theories directly
in field theory from the quotient viewpoint, which is one reason why
having a string realization is useful.

While the construction presented in \cite{Garcia-Etxebarria:2015wns}
using D3 branes is well suited to get the quotient of the $U(N)$
theory, it does not seem to generalize to the exceptional cases, since
there is no known construction of $\cN=4$ E-type SYM using D3
branes. However, as we will see in the following, one can obtain a
string realization of these by combining appropriately the
non-geometric construction of the (0,2) E-type theories presented
above with the $\cN=3$ quotient of section \ref{sec:sixdim}. Let us
start with M-theory on a five-torus,
$\bR^{1,3}\times S^1_a\times S^1_b\times S^1_c\times S^1_d\times
S^1_e\times \bC$,
where we will denote the different subtori as
$T^2_{ab}=S^1_a\times S^1_b$, etc.

\paragraph{E-type quotient.}  As explained in the previous section, the non-geometric orbifold that encodes the monodromies associated to the U-fold that yields the E-type (0,2) theories is $\bR^{1,3}\times T^2_{ab}\times (T^3_{cde}\times \bC)/\bZ^{E}_p$, where $\bZ^{E}_p$ acts as in \eqref{zetarho}, which we repeat here for convenience.
\begin{equation}\label{zetaE}
\bZ^E_p=\bZ_p^\bC\cdot \bZ_p^\rho\,,
\end{equation}
where $\bZ_p^\bC$ acts on $\bC$ as a rotation by $2\pi/p$ and $\bZ_p^\rho$ acts on the $\rho$ parameter of $T^3_{cde}$ (which we denote by $\rho_E$) by the monodromy in table \ref{table:ADE}. For $p=3,4,6$ we obtain the (0,2) $E_{6,7,8}$ theory on $\bR^{1,3}\times T^2_{ab}$. This quotient requires $\rho_E$ to be either $\rho_E=e^{i\pi/3}$ for $p=3,6$ or $\rho_E=i$ for $p=4$.

\paragraph{S-fold quotient.} The non-geometric quotient, discussed in section \ref{sec:sixdim}, that produces the S-fold is given by $\bR^{1,3}\times (T^3_{abc}\times T^2_{de}\times \bC)/\bZ^S_k$, where $\bZ^S_k$ is the combined action
\begin{equation}\label{zetaS}
\bZ^S_k=\bZ_k^R\cdot\tilde\bZ_k^R\cdot\bZ_k^\tau\,.
\end{equation} 
Here $\bZ_k^R$ is a rotation by $2\pi/k$ on $T^2_{de}\times \bC$, as in \eqref{rot3}, $\tilde \bZ_k^R$ acts on the $\rho$ parameter of $T^3_{abc}$ (denoted by $\rho_S$) via the monodromy in table \ref{table:ADE} and $\bZ_k^\tau$ is a $2\pi/k$ rotation on $T^2_{ab}$. In order to being able to perform this quotient, we need to set $\rho_S=\tau_{ab}=\tau_{de}=e^{i\pi/3}$ for $k=3,6$ and $\rho_S=\tau_{ab}=\tau_{de}=i$ for $k=4$.

Thus, we conclude that the four dimensional exceptional $\cN=3$ theories arise when we take the (non-geometric) S-fold quotient of the U-fold that produces the E-type (2,0) theories. In four dimensional field theory terms, the label $p$ denotes the parent gauge group $E_{6,7,8}$ and the label $k$ is the kind of $\cN=3$ quotient, or S-fold. If we take, for instance, the case $p=k=4$, the constraints of the $\rho$ and $\tau$ parameters imply that, out of the five independent radii of $T^5$, only one is independent, namely
\begin{equation}
R_a=R_b=R_d=R_e=R\,,\qquad R_c=R^{-2}\,.
\end{equation}
The four dimensional theory is reached when $R\rightarrow 0$.

In appendix \ref{app:details}, we describe how the monodromies are embedded in the duality group of M-theory on $T^5$, namely $O(5,5;\bZ)$. We use this embedding for computing the supercharges preserved by the non-geometric orbifold
\begin{equation}
\bR^{1,3}\times (\bC\times T^5)/(\bZ_p^E\times \bZ_k^S)\,.
\end{equation}
This computation shows that the theory indeed preserves twelve supercharges in four dimensions.

\section{Conclusions}
\label{sec:conclusions}

In this note we have constructed a new set of $\cN=3$ SCFTs in four
dimensions associated with exceptional algebras. We have no reason to
discard the existence of a purely geometric construction of these
theories, but the simplest approach, in terms of the $(0,2)$ $E$-type
theories in six-dimensions, led us naturally to consider M-theory on
U-manifolds, rather than an ordinary geometric compactification.

Along the way we have encountered a surprise: the action on the M5
brane worldvolume involves a discrete generator which is not part of
the usually considered (geometric) $SO(5)$ R-symmetry group. It should
be interesting to see whether use of this extra symmetry can be useful
for extending the class of $\cN=2$ theories in four dimensions that
can be analyzed along the lines of \cite{Gaiotto:2009we}. In
particular, it should be enlightening to approach, from this
six-dimensional viewpoint, the $\cN=3$ theories constructed here and
in \cite{Garcia-Etxebarria:2015wns}. A first interesting step would be
to reformulate the exotic discrete $\bZ_k$ action leading to the
$\cN=4 \to \cN=3$ breaking in terms of the two dimensional
theory associated with the four dimensional theory of interest
\cite{Alday:2009aq}. We hope to report on this topic in the future.

One well known aspect of $(0,2)$ theories compactified on $T^2$ is
that one can include variants with a non-trivial twisting by an outer
automorphism along the $T^2$, for instance in order to construct the
non-simply laced $\cN=4$ theories \cite{Vafa:1997mh,Tachikawa:2011ch},
or in order to construct new $\cN=2$ theories
\cite{Tachikawa:2010vg}. We have not considered twisting by such outer
automorphisms in this note, but it would be a very interesting
direction for further work. In particular, we expect such a
construction to describe some or all of the S-fold variants discussed
in \cite{Garcia-Etxebarria:2015wns,Aharony:2016kai}.

Another interesting direction is to generalize the class of
compactifications slightly, to include spaces in which the $T^2$
factor is fibered non-trivially over the four-dimensional base, while
preserving some supersymmetry. This setup appears naturally in the
$A_{N-1}$ case, where the resulting $\cN=4$ theories with duality
defects are useful for understanding aspects of the physics of
euclidean D3-branes in F-theory language
\cite{Witten:1996bn,Cvetic:2009ah,Blumenhagen:2010ja,Donagi:2010pd,Cvetic:2010ky,Grimm:2011dj,Marsano:2011nn,Cvetic:2011gp,Bianchi:2011qh,Bianchi:2012pn,Cvetic:2012ts,Bianchi:2012kt,Martucci:2014ema,Gadde:2014wma,Assel:2016wcr}. There
is no known corresponding notion of an ``exceptional instanton'', but
the abstract study of the generalization of such duality defects to
exceptional theories should be interesting in any case, and the
non-geometric backgrounds described in this note give one way of
explicitly constructing such setups.

More generally, it is natural to wonder whether non-geometric
engineering leads to a richer class of possibilities for constructing
six and four dimensional field theories beyond those accessible via
geometric techniques, as we suspect to be the case in the particular
example we have studied here. Of course, one always ends up having
to pay the piper: our understanding and control of U-manifolds,
particularly their moduli spaces and singularities, is still in its
infancy, so currently we can say rather less about this class of
theories than about those with geometric constructions. For instance,
a basic question about the theories we have constructed that is not
straightforward to answer is determining the dimension of their
Coulomb branch (for instance, in order to connect, down the road, with
the classification program of
\cite{Argyres:2015ffa,Argyres:2015gha,Argyres:2016xua,Argyres:2016xmc}).
The natural object to study would be the set of supersymmetric
deformations of the U-manifold we constructed, and to our knowledge
there is currently no simple way of approaching this question.

Nevertheless we hope that the existence of the exceptional theories
constructed in this note provides good motivation for taking this
``non-geometric engineering of QFTs'' viewpoint seriously, and
developing it further.

\acknowledgments

We thank Dieter L\"ust, Emanuel Malek and Daniel Park for enlightening
discussions. D.R thanks the University of Wisconsin-Madison for
hospitality.  D.R is supported by a grant from the Max Planck Society.

\appendix

\section{Details on the $\cN=3$ exceptional quotient}

\label{app:details}

The duality group of M-theory on $T^5$ is $O(5,5;\bZ)$. The maximal compact subgroup of the continuous version is $SO(5)\times SO(5)$, and the supercharges transform as
\begin{align}\label{one}
(S_6^+,\mathbf{1},\mathbf{4})\oplus(S_6^-,\mathbf{4},\mathbf{1})
\end{align}
under $SO(1,5)\times SO(5)\times SO(5)$. We would like to understand how the monodromies of the E-type and S-fold quotients are embedded in $SO(5)\times SO(5)$ and how they act on the supercharges. Recall that the $\mathbf 4$ of $SO(5)$ is
\begin{align}
\left \{\left (\frac12\,\frac12\right ),\,\left (\frac12\,-\frac12\right ),\,\left (-\frac12\,\frac12\right ),\,\left (-\frac12\,-\frac12\right )\right \}\,.
\end{align}
In the following we will omit the $\frac12$ and simply write $(++)$, etc.

Firstly, there is an $SO(5)_S\subset SO(5)\times SO(5)$ that corresponds to the structure group of $T^5$, i.e.~the geometrical $SO(5)_S$. Clearly, such a subgroup must act in the same way on the supercharges of both chiralities. This can be seen by taking an eleven dimensional spinor and decomposing it according to the splitting $\bR^{1,10}\rightarrow \bR^{1,5}\times T^5$, namely
\begin{align}\label{two}
S_{11}=(S_6^+,\mathbf 4)\oplus(S_6^-,\mathbf 4)\,.
\end{align}
Thus, from \eqref{one} and \eqref{two}, we see that $SO(5)_S$ is embedded as
\begin{align}
SO(5)_S=\{(g,g)\in SO(5)\times SO(5)\}\,,
\end{align}
so it is the diagonal subgroup. We define also the anti-diagonal subgroup
\begin{align}\label{A}
SO(5)_A=\{(g,g^{-1})\in SO(5)\times SO(5)\}\,.
\end{align}

The E-type and S-fold quotients involve geometric rotations in two different tori, $T^{ab}$ and $T^{de}$, of $T^5$. Then, we choose the Cartan subalgebra of $SO(5)\times SO(5)$ such that the Cartan of $SO(5)_S$ corresponds to rotations along $T^2_{ab}$ and $T^2_{de}$. The quotients also involve non-geometric actions on two three-tori, $T^3_{abc}$ and $T^3_{cde}$, which correspond to the Cartan of $SO(5)_A$.

\paragraph{Example: Geometric rotation around $T^2_{ab}$.} Consider a rotation of $2\pi/k$ around $T^2_{ab}$. The action on the supercharges \eqref{one} is
\begin{align}
\left ( S_6^+,\mathbf 1,\begin{array}{c}++\\+-\\-+\\--\end{array}\right )&\rightarrow \exp\left \{{2\pi i\frac{1}{2k}\left (\begin{array}{c} +1\\+1\\-1\\-1 \end{array}\right )}\right \}\left ( S_6^+,\mathbf 1,\begin{array}{c}++\\+-\\-+\\--\end{array}\right )\\
\left ( S_6^-,\begin{array}{c}++\\+-\\-+\\--\end{array},\mathbf 1\right )&\rightarrow \exp\left \{{2\pi i\frac{1}{2k}\left (\begin{array}{c} +1\\+1\\-1\\-1 \end{array}\right )}\right \}\left ( S_6^+,\begin{array}{c}++\\+-\\-+\\--\end{array},\mathbf 1\right )\,.
\end{align}
The rotation acts on the first entry of the weight and in the same way for both supercharges.

\paragraph{Example: Non-geometric rotation around $T^2_{abc}$.} Consider a non-geometric rotation of $2\pi/k$ around $T^3_{abc}$. The action on the supercharges \eqref{one} is dictated by the embedding \eqref{A} and is
\begin{align}
\left ( S_6^+,\mathbf 1,\begin{array}{c}++\\+-\\-+\\--\end{array}\right )&\rightarrow \exp\left \{{2\pi i\frac{1}{2k}\left (\begin{array}{c} -1\\-1\\+1\\+1 \end{array}\right )}\right \}\left ( S_6^+,\mathbf 1,\begin{array}{c}++\\+-\\-+\\--\end{array}\right )\\
\left ( S_6^-,\begin{array}{c}++\\+-\\-+\\--\end{array},\mathbf 1\right )&\rightarrow \exp\left \{{2\pi i\frac{1}{2k}\left (\begin{array}{c} +1\\+1\\-1\\-1 \end{array}\right )}\right \}\left ( S_6^+,\begin{array}{c}++\\+-\\-+\\--\end{array},\mathbf 1\right )\,.
\end{align}
The non-geometric rotation acts on the first entry of the weight and in the opposite way for both supercharges.

\subsection*{Supercharges}

The quotient that yields the exceptional $\cN=3$ theories is $\bZ_p^E\times \bZ^S_k$, so the supercharges that survive must be invariant under both $\bZ_p^E$ and $\bZ_k^S$ separately. Since these quotients involve an additional $\bC$, we split the supercharges in \eqref{one} as
\begin{align}\label{three}
(S_4^+,\frac12,\mathbf{1},\mathbf{4})\oplus(S_4^-,-\frac12,\mathbf{1},\mathbf{4})\oplus(S_4^+,-\frac12,\mathbf{4},\mathbf{1})\oplus(S_4^-,+\frac12,\mathbf{4},\mathbf{1})\,,
\end{align}
which is how they transform under $SO(1,3)\times U(1)_\bC\times SO(5)\times SO(5)$.

\paragraph{E-type quotient.} This consists of a rotation around $\bC$ together with a non-geometric rotation around $T^3_{cde}$. The latter acts on the second weight in the opposite way for $(\mathbf 1, \mathbf 4)$ and $(\mathbf 4, \mathbf 1)$.

\noindent We find
\begin{align}
\left ( S_4^+,\frac12,\mathbf 1,\begin{array}{c}++\\+-\\-+\\--\end{array}\right )&\rightarrow \exp\left \{{2\pi i\frac{1}{2k}\left (\begin{array}{c} 1-1\\1+1\\1-1\\1+1 \end{array}\right )}\right \}\left ( S_4^+,\frac12,\mathbf 1,\begin{array}{c}++\\+-\\-+\\--\end{array}\right )\\
\left ( S_4^-,-\frac12,\mathbf 1,\begin{array}{c}++\\+-\\-+\\--\end{array}\right )&\rightarrow \exp\left \{{2\pi i\frac{1}{2k}\left (\begin{array}{c} -1-1\\-1+1\\-1-1\\-1+1 \end{array}\right )}\right \}\left ( S_4^-,-\frac12,\mathbf 1,\begin{array}{c}++\\+-\\-+\\--\end{array}\right )\\
\left ( S_4^+,-\frac12,\begin{array}{c}++\\+-\\-+\\--\end{array},\mathbf 1\right )&\rightarrow \exp\left \{{2\pi i\frac{1}{2k}\left (\begin{array}{c} -1+1\\-1-1\\-1+1\\-1-1 \end{array}\right )}\right \}\left ( S_4^+,-\frac12,\begin{array}{c}++\\+-\\-+\\--\end{array},\mathbf 1\right )\\
\left ( S_4^-,\frac12,\begin{array}{c}++\\+-\\-+\\--\end{array},\mathbf 1\right )&\rightarrow \exp\left \{{2\pi i\frac{1}{2k}\left (\begin{array}{c} 1+1\\1-1\\1+1\\1-1 \end{array}\right )}\right \}\left ( S_4^-,\frac12,\begin{array}{c}++\\+-\\-+\\--\end{array},\mathbf 1\right )\,.
\end{align}
This shows that only half of the supercharges survive, namely
\begin{align}\label{Etype}
\begin{split}
\left ( S_4^+,\frac12,\mathbf 1,++\right )\oplus\left ( S_4^+,\frac12,\mathbf 1,-+\right )\oplus\left ( S_4^-,-\frac12,\mathbf 1,+-\right )\oplus\left ( S_4^-,-\frac12,\mathbf 1,--\right )\oplus\\
\left ( S_4^+,-\frac12,++,\mathbf 1\right )\oplus\left ( S_4^+,-\frac12,-+,\mathbf 1\right )\oplus\left ( S_4^-,\frac12,+-,\mathbf 1\right )\oplus\left ( S_4^-,\frac12,--,\mathbf 1\right )\,.
\end{split}
\end{align}

\paragraph{S-fold quotient.} In this case we have to take a quotient by a geometric rotation in $\bC$, $T^2_{ab}$ and $T^2_{de}$. The last two act on the first and second entries of the weight vector, respectively, and in the same way for $(\mathbf 1, \mathbf 4)$ and $(\mathbf 4, \mathbf 1)$. We also need to take a quotient by a non-geometric rotation in $T^3_{abc}$. 

The action of this quotient on the supercharges \eqref{Etype} that survive the E-type quotient is
\begin{align}
\left ( S_4^+,\frac12,\mathbf 1,++\right )&\rightarrow \exp\left \{ 2\pi i\frac{1}{2k}(1+1-1-1) \right \} \left ( S_4^+,\frac12,\mathbf 1,++\right )\,.
\end{align}
The four contributions to the phase $\exp\left \{ 2\pi i\frac{1}{2k}(1+1-1-1) \right \}$ should be understood as coming, in order, from: 1) rotation in $\bC$, 2) rotation in $T^2_{ab}$, 3) rotation in $T^2_{de}$ and 4) non-geometric rotation in $T^3_{abc}$. The action on the rest is
\begin{align}
\left ( S_4^+,\frac12,\mathbf 1,-+\right )&\rightarrow \exp\left \{ 2\pi i\frac{1}{2k}(1-1-1+1) \right \}\left ( S_4^+,\frac12,\mathbf 1,-+\right )\\
\left ( S_4^-,-\frac12,\mathbf 1,+-\right )&\rightarrow \exp\left \{ 2\pi i\frac{1}{2k}(-1+1+1-1) \right \} \left ( S_4^-,-\frac12,\mathbf 1,+-\right )\\
\left ( S_4^-,-\frac12,\mathbf 1,--\right )&\rightarrow \exp\left \{ 2\pi i\frac{1}{2k}(-1-1+1+1) \right \} \left ( S_4^-,-\frac12,\mathbf 1,--\right )\\
\left ( S_4^+,-\frac12,++,\mathbf 1\right )&\rightarrow \exp\left \{ 2\pi i\frac{1}{2k}(-1+1-1+1) \right \} \left ( S_4^+,-\frac12,++,\mathbf 1\right )\\
\left ( S_4^+,-\frac12,-+,\mathbf 1\right )&\rightarrow \exp\left \{ 2\pi i\frac{1}{2k}(-1-1-1-1) \right \} \left ( S_4^+,-\frac12,-+,\mathbf 1\right )\\
\left ( S_4^-,\frac12,+-,\mathbf 1\right )&\rightarrow \exp\left \{ 2\pi i\frac{1}{2k}(1+1+1+1) \right \}\left ( S_4^-,\frac12,+-,\mathbf 1\right )\\
\left ( S_4^-,\frac12,--,\mathbf 1\right )&\rightarrow \exp\left \{ 2\pi i\frac{1}{2k}(1-1+1-1) \right \}\left ( S_4^-,\frac12,--,\mathbf 1\right )\,.
\end{align}
We see that, in total, twelve supercharges survive the action of $\bZ_p^E$ and $\bZ_k^S$. Thus, the quotient $\bZ_p^E\times \bZ^S_k$ yields an $\cN=3$ theory in four dimensions.

\bibliographystyle{JHEP}
\bibliography{refs}

\end{document}